\documentclass[aps,prl,preprint,showpacs,showkeys]{revtex4}
\usepackage{amsmath}
\usepackage{graphicx}
\usepackage{bm}
\input{tcilatex}

\begin{document}

\title{Remarks on the Solution of the Position Dependent Mass (PDM) Schr\"{o}%
dinger Equation}
\author{Ramazan Ko\c{c} and Seda Say\i n}
\affiliation{Gaziantep University, Department of Physics, Faculty of Engineering 27310
Gaziantep/Turkey}
\email{koc@gantep.edu.tr, ssayin@gantep.edu.tr}
\date{\today }

\begin{abstract}
An approximate method is proposed to solve position dependent mass Schr\"{o}%
dinger equation. The procedure suggested here leads to the solution of the
PDM Schr\"{o}dinger equation without transforming the potential function to
the mass space or vice verse. The method based on asymptotic Taylor
expansion of the function, produces an approximate analytical expression for
eigenfunction and numerical results for eigenvalues of the PDM Schr\"{o}%
dinger equation. The results show that PDM and constant mass Schr\"{o}dinger
equations are not isospectral. The calculations are carried out with the aid
of a computer system of symbolic or numerical calculation by constructing a
simple algorithm.
\end{abstract}

\keywords{Position dependent mass Schr\"{o}dinger equation, Taylor
Expansion, Function theory, Numerical approximation and analysis }
\pacs{03.65.Ge, 03.65.Fd, 02.30.-f, 02.60.-x }
\maketitle

\section{Introduction}

Quantum mechanical systems with a position dependent mass (PDM) generate
interest for its relevance and importance in describing the physics of many
microstructures of current interest, understanding transport phenomena in
compositionally graded crystals, designing modern fabrication of nano
devices such as quantum dots, wires and wells, developing theoretical models
for effective interactions in nuclear physics, neutron stars, liquid
crystals, metal clusters\cite%
{Serra,Barranco,Bastard,Koc0,Geller,Arias,Harrison,Weis,Peter,Khor,Alex,Taro,Fu,Young,von1}%
. These applications have stimulated a naturally renewed interest in the
solution of PDM quantum mechanical Hamiltonians. Recently, solution of the
PDM Schr\"{o}dinger equation, Dirac equation \cite{Haidari,Roy,Ikdar1,Peng}
and Klein-Gordon equation \cite{Ikdar2,Arda,Dutra} have received much
attention. A number of authors have studied PDM Schr\"{o}dinger equation
within the framework of point canonical transformations \cite%
{Quesne1,Tezcan,Chen,Mustafa}, Lie algebraic techniques\cite%
{Roy2,Kerimov,Yahio,Quesne2,Dong,Roy3,Jana,Cruz}, super symmetric
quantum-mechanical \cite%
{Quesne3,Tanaka,Suzko,Koc1,Bagchi,Ganguly,Gonul,Bagchi1,Cannata}, or other
related techniques \cite%
{Roy4,Koc2,Koc3,Gang,Chen2,Bagchi2,Haidari2,Jiang,Ioffe,Kuru,Carin,Midya,Qu,Chen3,Gorain,Ju,Aktas,Sever,Kraen,Dutta}%
.

In most applications of such methods, PDM Schr\"{o}dinger equation has been
transformed in the form of the constant mass Schr\"{o}dinger equation by
changing coordinate and wave function. Obviously, this transformation
generates isospectral potentials and exact solvability requirements result
in constraints on the potential functions for the given mass distributions.
In other words, a suitable transformation of coordinate and wave function
becomes a bridge between constant mass and position dependent mass Schr\"{o}%
dinger equation. As an example in a constant mass Schr\"{o}dinger equation
the choice of coordinate $u=\int_{0}^{x}\sqrt{m(x)}dx$ and wave function $%
\psi (u)=\left[ 2m(x)\right] ^{1/4}\varphi (x)$ provides its transformation
in the form of the PDM Schr\"{o}dinger equation. In this case the potential
is mass dependent; \textit{i. e}. harmonic oscillator potential can be
expressed as $V=\frac{1}{2}m\omega ^{2}u^{2}=\frac{1}{2}m\omega ^{2}\left(
\int_{0}^{x}\sqrt{m(x)}dx\right) ^{2}$ and both constant and PDM Schr\"{o}%
dinger equations have the same eigenvalues. The origin of such an
isospectrality in the constant mass scenario has not yet been studied. It
will be worthwhile to discuss physical acceptability of such an
isospectrality in the position dependent mass background. In some articles 
\cite{Peter,Khor,Alex,Taro,Arda,Gang,Kuru,Chen3,Sever}, solution of the PDM
Schr\"{o}dinger equation has been obtained without transforming the
potential in to mass space . In this case the energy spectrum of the PDM
Hamiltonians are not isospectral with the constant mass Hamiltonians.
Therefore it is reasonable to develop a method for solving PDM Hamiltonian
without transforming the potential into the mass space.

However, the fundamental question remains open: how the potential is
affected when it is expressed in the mass space? To answer this question,
one has to obtain a solution for the Schr\"{o}dinger equation without
transforming the potential to the mass space. In this article, we will
obtain a semi-analytical solution of the Schr\"{o}dinger equation without
transforming the potential to the mass space. This is another reason, to
build a realistic model for solving PDM Hamiltonian.

It is well known that the study of the same mathematical problems from
different point of view lead to the progress of the science and includes a
lot of mathematical tastes. A technique based on asymptotic expansion of
Taylor series \cite{Koc4} has recently been suggested to obtain eigenvalues
of Schr\"{o}dinger equation which improves both analytical and numerical
determination of the eigenvalues. Asymptotic Taylor Expansion Method (ATEM)
is very efficient to obtain eigenvalues of the Schr\"{o}dinger equation
because of their simplicity and low round off error. This method has been
easily applied to establish eigenvalues and wave function of the Schr\"{o}%
dinger type equations. We would like to mention here that the ATEM is a
field of tremendous scope and has an almost unlimited opportunity, for its
applications in the solution of the constant and PDM Schr\"{o}dinger
equations. In this paper, we address ourselves to the solution of the PDM
Schr\"{o}dinger equation by using the ATEM.

The paper is organized as follows. In the next section we review
construction of ATEM by reformulating the well known Taylor series expansion
of a function that satisfies second order homogeneous differential equation
of the form: $f^{\prime \prime }(x)=p_{0}(x)f^{\prime }(x)+q_{0}(x)f(x)$.
Section 3 is devoted to the application of the main result for solving the
PDM Schr\"{o}dinger equation for various forms of the Kinetic energy
operator. As a practical example, we illustrate solution of the PDM Schr\"{o}%
dinger equation including harmonic oscillator potential and variable mass $%
m(x)=m_{0}\left( 1+\gamma x^{2}\right) $ \cite{Alex}. In this section, we
present an approximate analytical expression for eigenfunction and numerical
results for eigenvalues of the PDM Schr\"{o}dinger equation. We also analyze
the asymptotic behavior of the Hamiltonian. Some concluding remarks are
given in section 4.

\section{Formalism of ATEM}

In this section, we show the solution of the Schr\"{o}dinger type equation
for a quite ample class of potentials, by modifying Taylor series expansion
by means of a finite sequence instead of an infinite sequence and its
termination possessing the property of quantum mechanical wave function. Let
us consider Taylor series expansion \cite{Taylor} of a function $f(x)$ about
the point $a$:%
\begin{eqnarray}
f(x) &=&f(a)+(x-a)f^{\prime }(a)+\frac{1}{2}(x-a)^{2}f^{\prime \prime }(a)+%
\frac{1}{6}(x-a)^{3}f^{(3)}(a)+........  \notag \\
&&\overset{\infty }{=\underset{n=0}{\dsum }}\frac{(x-a)^{n}}{n!}f^{(n)}(a)
\label{a1}
\end{eqnarray}%
where $f^{(n)}(a)$ is the $n^{th}$ derivative of the function at $a$. Taylor
series specifies the value of a function at one point, $x$, in terms of the
value of the function and its derivatives at a reference point $a$.
Expansion of the function $f(x)$ about the origin ($a=0$), is known as
Maclaurin's series and it is given by,%
\begin{eqnarray}
f(x) &=&f(0)+xf^{\prime }(0)+\frac{1}{2}x^{2}f^{\prime \prime }(0)+\frac{1}{6%
}x^{3}f^{(3)}(0)+........  \notag \\
&&\overset{\infty }{=\underset{n=0}{\dsum }}\frac{x^{n}}{n!}f^{(n)}(0).
\label{a11}
\end{eqnarray}%
Here we develop a method to solve a second order linear differential
equation of the form: 
\begin{equation}
f^{\prime \prime }(x)=p_{0}(x)f^{\prime }(x)+q_{0}(x)f(x).  \label{a2}
\end{equation}%
It is obvious that the higher order derivatives of the $f(x)$ can be
obtained in terms of the $f(x)$ and $f^{\prime }(x)$ by differentiating (\ref%
{a2}). Then, higher order derivatives of $f(x)$ are given by%
\begin{equation}
f^{(n+2)}(x)=p_{n}(x)f^{\prime }(x)+q_{n}(x)f(x)  \label{a3}
\end{equation}%
where%
\begin{eqnarray}
p_{n}(x) &=&p_{0}(x)p_{n-1}(x)+p_{n-1}^{\prime }(x)+q_{n-1}(x),\text{ and} 
\notag \\
q_{n}(x) &=&q_{0}(x)p_{n-1}(x)+q_{n-1}^{\prime }(x).  \label{a4}
\end{eqnarray}%
Of course, the last result shows there exist a formal relation between
asymptotic iteration method (AIM) \cite{ciftci} and ATEM. We have observed
that eigenfunction of the \ Schr\"{o}dinger type equations can efficiently
be determined by using ATEM. It is clear that the recurrence relations (\ref%
{a4}) allow us algebraic exact or approximate analytical expression for the
solution of (\ref{a2}) under some certain conditions. Let us substitute (\ref%
{a4}) into the (\ref{a1}) to obtain%
\begin{equation}
f(x)=f(0)\left( 1+\overset{m}{\underset{n=2}{\dsum }}q_{n-2}(0)\frac{x^{n}}{%
n!}\right) +f^{\prime }(0)\left( 1+\overset{m}{\underset{n=2}{\dsum }}%
p_{n-2}(0)\frac{x^{n}}{n!}\right) .  \label{a5}
\end{equation}%
After all, we have obtained useful formalism of the Taylor expansion method.
In the solution of the eigenvalue problems, truncation of the the asymptotic
expansion to a finite number of terms is useful. If the series optimally
truncated at the smallest term then the asymptotic expansion of series is
known as superasymptotic \cite{boyd}, and it leads to the determination of
eigenvalues with minimum error. Then boundary conditions can be applied as
follows. When only odd or even power of $x$ collected as coefficients of $%
f(0)$ or $f^{\prime }(0)$ and vice verse, the series is truncated at $n=m$
then an immediate practical consequence of these condition for $q_{m-2}(0)=0$
or $p_{m-2}(0)=0.$ In this way, one of the parameter in the $q_{m-2}(0)$
and/or $p_{m-2}(0)$ belongs to the spectrum of the Schr\"{o}dinger equation.
Therefore eigenfunction of the equation becomes a polynomial of degree $m$.
Otherwise the spectrum of the system can be obtained as follows: In a
quantum mechanical system eigenfunction of the system is discrete. Therefore
in order to terminate the eigenfunction $f(x)$ we can concisely write that%
\begin{eqnarray}
q_{m}(0)f(0)+p_{m}(0)f^{\prime }(0) &=&0  \notag \\
q_{m-1}(0)f(0)+p_{m-1}(0)f^{\prime }(0) &=&0  \label{a5y}
\end{eqnarray}%
eliminating $f(0)$ and $f^{\prime }(0)$ we obtain%
\begin{equation}
q_{m}(0)p_{m-1}(0)-p_{m}(0)q_{m-1}(0)=0  \label{a5z}
\end{equation}%
again one of the parameter in the equation related to the eigenvalues of the
problem.

In quantum mechanics bound state energy of the atom is quantized and
eigenvalues are discrete and for each eigenvalues there exist one or more
eigenfunctions. When we are dealing with the solution of the Schr\"{o}dinger
equation, we are mainly interested in the discrete eigenvalues of the
problem. The first main result of this conclusion gives necessary and
sufficient conditions for the termination of the Taylor series expansion of
the wave function.

The process presented here is iterative and number of iteration is given by $%
m$. The results are obtained as follows: in our Mathematica program, we use
an iteration number, say $m=30$, then we obtain another result for $m=40$,
so on, then we compare values of the parameter (eigenvalue) in each case
till $10$ digits. If values of the parameter reach its asymptotic value then
we use these values and omit the others. For instance, if one can obtain
values of the parameters for $m=40$, first few of them will be reached its
asymptotic values, say first $8$ values. The following comment for the
function is considerable: for such a solution it is suitable to take sum of
first $8$ term in the (\ref{a5}).

It will be shown that ATEM gives accurate results for PDM Schr\"{o}dinger
equations. In the following sections, it is shown that this approach opens
the way to the treatment of \ PDM\ Schr\"{o}dinger equation including large
class of potentials of practical interest.

\section{Solution of the PDM Schr\"{o}dinger equation by using ATEM}

In the PDM Schr\"{o}dinger equation the mass and momentum operator no longer
commute, so there are several ways to define kinetic energy operator. The
general expression for the Hamiltonian with the kinetic energy operator
introduced by von Roos\cite{von2} and potential energy $V(x)$, can be
written as:%
\begin{equation}
H=\frac{1}{4}\left( m^{\eta }\mathbf{p}m^{\varepsilon }\mathbf{p}m^{\rho
}+m^{\rho }\mathbf{p}m^{\varepsilon }\mathbf{p}m^{\eta }\right) +V(x)
\label{eq:1}
\end{equation}%
where $\eta +\varepsilon +\rho =-1$ is a constraint and $m=m(x)$ is position
dependent mass. There are many debates for the choice of the parameters $%
\eta ,$ $\varepsilon ,$and $\rho $, in our approach, we will obtain the
solution of the PDM Schr\"{o}dinger equation for the following Hamiltonians 
\cite{von2,Dutra2,Dekar}: 
\begin{subequations}
\begin{eqnarray}
H_{1} &=&\frac{1}{2}\left( \mathbf{p}\frac{1}{m}\mathbf{p}\right) +V(x);%
\text{ for }\varepsilon =-1,\rho =0,\eta =0,  \label{eq:2a} \\
H_{2} &=&\frac{1}{4}\left( \frac{1}{m}\mathbf{p}^{2}+\mathbf{p}^{2}\frac{1}{m%
}\right) +V(x);\text{ for }\varepsilon =0,\rho =0,\eta =-1,  \label{eq:2b} \\
H_{3} &=&\frac{1}{2}\left( \frac{1}{\sqrt{m}}\mathbf{p}^{2}\frac{1}{\sqrt{m}}%
\right) +V(x);\text{ for }\varepsilon =-\frac{1}{2},\rho =0,\eta =-\frac{1}{2%
},  \label{eq:2c} \\
H_{4} &=&\frac{1}{2}\left( \mathbf{p}\frac{1}{\sqrt{m}}\mathbf{p}\frac{1}{%
\sqrt{m}}+\frac{1}{\sqrt{m}}\mathbf{p}\frac{1}{\sqrt{m}}\mathbf{p}\right)
+V(x);\text{ for }\varepsilon =0,\rho =-\frac{1}{2},\eta =-\frac{1}{2}.
\label{eq:2d}
\end{eqnarray}%
Here we take a new look at the solution of the the PDM Schr\"{o}dinger
equation by using the method of ATEM developed in the previous section.

Before going further we share one of our significant observation during our
calculations. If the mass distribution is not appropriate for a given
potential, the eigenvalues do not reach their asymptotic values and
resultant eigenfunction cannot be terminated when $x\rightarrow \pm \infty $%
. In order to illustrate semi analytical solution of the eigenvalue
equations 
\end{subequations}
\begin{equation}
H_{i}\psi (x)=E\psi (x),(i=1,2,3,4)  \label{eq:3}
\end{equation}%
including harmonic oscillator potential, $V(x)=\frac{1}{2}m_{0}\omega
^{2}x^{2}$, we use the mass distributions $m(x)=m_{0}\left( 1+\gamma
x^{2}\right) $, where $\gamma $ is arbitrary positive constant. By the way,
we emphasize that the wave function of harmonic oscillator potential is well
defined in the region of $\pm \infty $ and satisfy that $\underset{%
x\rightarrow \pm \infty }{\lim }\frac{\left\vert \psi (x)\right\vert ^{2}}{%
\sqrt{m}}\rightarrow 0.$ In this limit the mass distributions to be
continuous.

It is well known that asymptotic behavior of constant mass Schr\"{o}dinger
equation including harmonic oscillator potential is given by $\psi =e^{-%
\frac{x^{2}}{2}}f(x),$ for simplicity we set $\hbar =m_{0}=\omega =1.$Thus,
this change of wave function guaranties $\underset{x\rightarrow \pm \infty }{%
\lim }\frac{\left\vert \psi (x)\right\vert ^{2}}{\sqrt{m}}\rightarrow 0$.
After this transformation, we present an iteration algorithm to calculate
both eigenvalues and eigenfunctions of the eigenvalue equation (\ref{eq:3}).
Using this algorithm, we develop a Mathematica program, which demonstrates
that it is easier to be implemented into a computer program, and produces a
highly accurate solution with analytical expression efficiently.

\subsubsection{Asymptotic Analysis}

The term asymptotic means the function approaching to a given value as the
iteration number tends to infinity. By the aid of a Mathematica program we
calculate eigenvalues and eigenfunction of $H_{1}$ for $\gamma =0.1$\ using
number of iterations $k=\{20,30,40,50$,$60\}$. The function $f(x)$ for $n=2$
state is given in (\ref{eq:4}) and eigenvalues are presented in Table I.

\begin{eqnarray}
k=20;\text{ }f(x)= &&1-1.857x^{2}-1.619\times 10^{-1}x^{4}+2.060\times
10^{-2}x^{6}  \notag \\
&&+1.515\times 10^{-3}x^{8}-1.261\times 10^{-4}x^{10}-6.495\times
10^{-6}x^{12}  \notag \\
k=40;\text{ }f(x)= &&1-1.856x^{2}-1.622\times 10^{-1}x^{4}+2.051\times
10^{-2}x^{6}  \notag \\
&&+1.505\times 10^{-3}x^{8}-1.271\times 10^{-4}x^{10}-6.662\times
10^{-6}x^{12}  \label{eq:4} \\
k=60;\text{ }f(x)= &&1-1.856x^{2}-1.622\times 10^{-1}x^{4}+2.051\times
10^{-2}x^{6}  \notag \\
&&+1.504\times 10^{-3}x^{8}-1.271\times 10^{-4}x^{10}-6.663\times
10^{-6}x^{12}  \notag
\end{eqnarray}%
Our calculation gives an accurate result for first $8$ eigenvalues and
eigenfunctions after $40$ iterations. Here we have used $60$ iterations.
Figure 1 shows the plot of normalized wave functions for first $6$ state. 
\begin{table}[tbph]
\begin{tabular}{|l|l|l|l|l|l|l|}
\hline
$k$ & $n=0$ & $n=1$ & $n=2$ & $n=3$ & $n=4$ & $n=5$ \\ \hline
$20$ & $0.46889047$ & $1.43341211$ & $2.35765542$ & $3.28397486$ & $%
4.21360362$ & $4.35399596$ \\ \hline
$30$ & $0.46889665$ & $1.43348058$ & $2.35642259$ & $3.24660834$ & $%
4.12086916$ & $4.98321327$ \\ \hline
$40$ & $0.46889650$ & $1.43348582$ & $2.35655507$ & $3.24585555$ & $%
4.10543833$ & $4.95755341$ \\ \hline
$50$ & $0.46889651$ & $1.43348553$ & $2.35654885$ & $3.24599291$ & $%
4.10703835$ & $4.94114551$ \\ \hline
$60$ & $0.46889651$ & $1.43348555$ & $2.35654908$ & $3.24598255$ & $%
4.10694346$ & $4.94337909$ \\ \hline
\end{tabular}%
\caption[The eigenvalues of quantum dot]{Eigenvalues of the PDM $H_{1}$ for
different iteration numbers $k$ and $\protect\gamma =0.1$.}
\end{table}

\begin{figure}[tbp]
\caption{Plot of the normalized wave function of the PDM Hamiltonian (%
\protect\ref{eq:2a}) for $n=0,1,2,3,4,5$. }
\QTP{Dialog Text}
\begin{tabular}{ll}
\FRAME{itbpF}{3.3797in}{2.3255in}{0in}{}{}{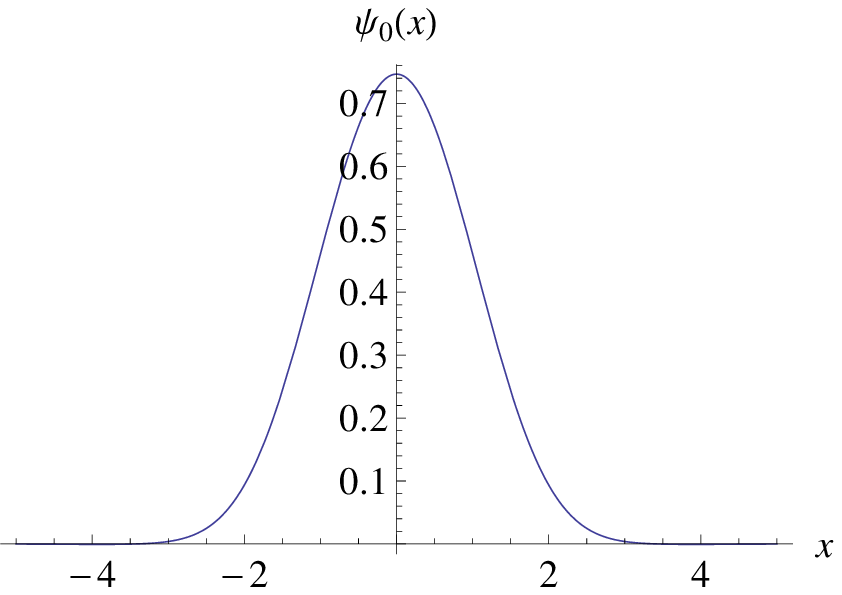}{\special{language
"Scientific Word";type "GRAPHIC";maintain-aspect-ratio TRUE;display
"USEDEF";valid_file "F";width 3.3797in;height 2.3255in;depth
0in;original-width 3.333in;original-height 2.2857in;cropleft "0";croptop
"1";cropright "1";cropbottom "0";filename 'fig0.eps';file-properties
"XNPEU";}} & \FRAME{itbpF}{3.3797in}{2.2157in}{0in}{}{}{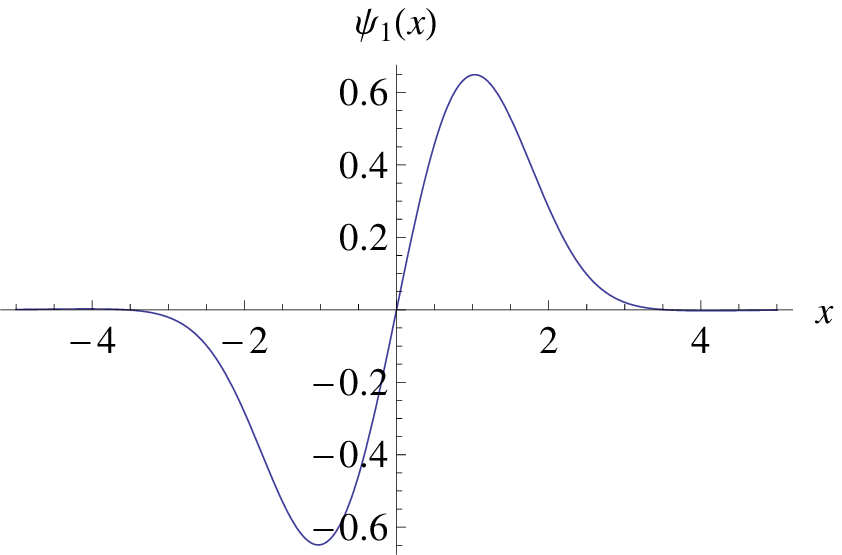}{\special%
{language "Scientific Word";type "GRAPHIC";maintain-aspect-ratio
TRUE;display "USEDEF";valid_file "F";width 3.3797in;height 2.2157in;depth
0in;original-width 3.333in;original-height 2.175in;cropleft "0";croptop
"1";cropright "1";cropbottom "0";filename 'fig1.eps';file-properties
"XNPEU";}} \\ 
\FRAME{itbpF}{3.3797in}{2.2157in}{0in}{}{}{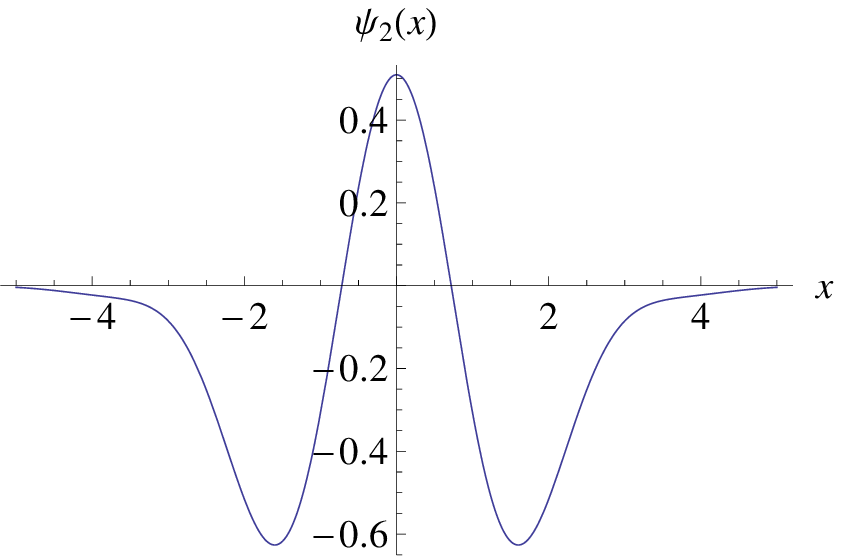}{\special{language
"Scientific Word";type "GRAPHIC";maintain-aspect-ratio TRUE;display
"USEDEF";valid_file "F";width 3.3797in;height 2.2157in;depth
0in;original-width 3.333in;original-height 2.175in;cropleft "0";croptop
"1";cropright "1";cropbottom "0";filename 'fig2.eps';file-properties
"XNPEU";}} & \FRAME{itbpF}{3.3797in}{2.2157in}{0in}{}{}{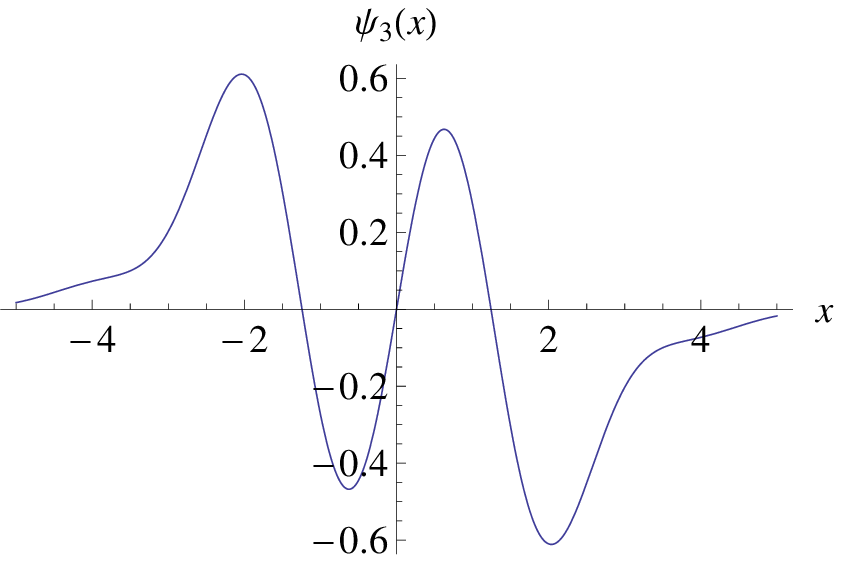}{\special%
{language "Scientific Word";type "GRAPHIC";maintain-aspect-ratio
TRUE;display "USEDEF";valid_file "F";width 3.3797in;height 2.2157in;depth
0in;original-width 3.333in;original-height 2.175in;cropleft "0";croptop
"1";cropright "1";cropbottom "0";filename 'fig3.eps';file-properties
"XNPEU";}} \\ 
\FRAME{itbpF}{3.3797in}{2.2157in}{0in}{}{}{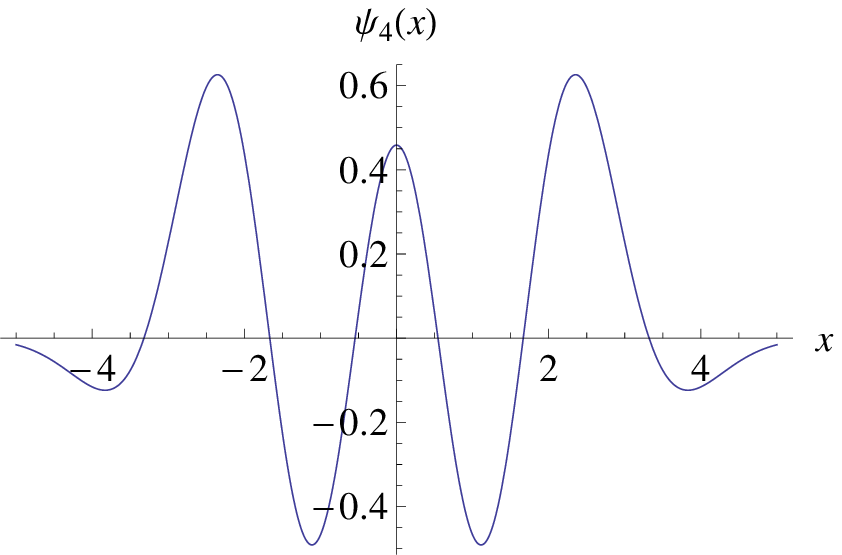}{\special{language
"Scientific Word";type "GRAPHIC";maintain-aspect-ratio TRUE;display
"USEDEF";valid_file "F";width 3.3797in;height 2.2157in;depth
0in;original-width 3.333in;original-height 2.175in;cropleft "0";croptop
"1";cropright "1";cropbottom "0";filename 'fig4.eps';file-properties
"XNPEU";}} & \FRAME{itbpF}{3.3797in}{2.2157in}{0in}{}{}{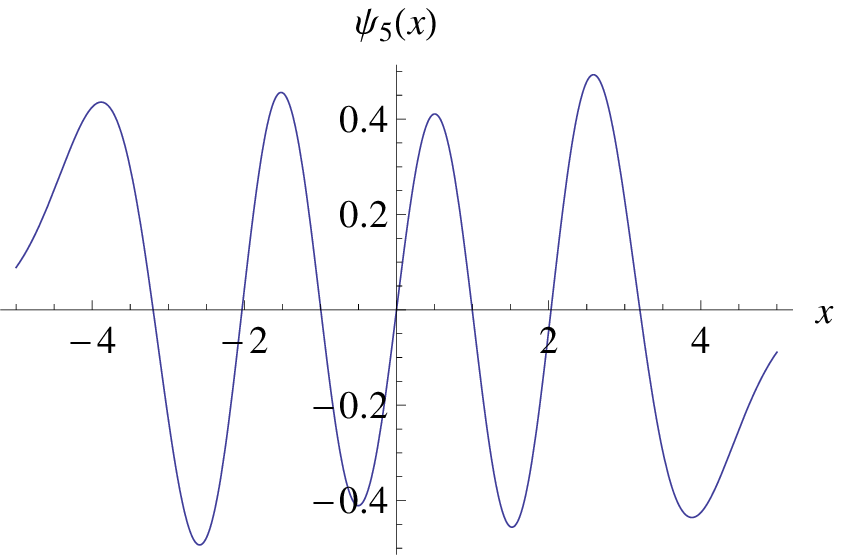}{\special%
{language "Scientific Word";type "GRAPHIC";maintain-aspect-ratio
TRUE;display "USEDEF";valid_file "F";width 3.3797in;height 2.2157in;depth
0in;original-width 3.333in;original-height 2.175in;cropleft "0";croptop
"1";cropright "1";cropbottom "0";filename 'fig5.eps';file-properties
"XNPEU";}}%
\end{tabular}%
\end{figure}

\subsubsection{Solution of the Hamiltonians $H_{2},$ $H_{3}$ and $H_{4}$}

In the previous section we have illustrated applicability of our method by
solving Hamiltonian $H_{1}.$ In this section we apply the same procedure to
solve the Hamiltonians $H_{2},$ $H_{3}$ and $H_{4}.$Again we have used $60$
iterations for each Hamiltonians and checked stability of the eigenvalues.
Here we calculated eigenvalues for $30$ iterations and they are listed in
Table II. We have also checked that for the given eigenvalues, the wave
functions are normalizable and it tends to zero when $x\rightarrow \infty $. 
\begin{table}[tbph]
\begin{tabular}{|l|l|l|l|}
\hline
$n$ & $E_{H_{2}}$ & $E_{H_{3}}$ & $E_{H_{4}}$ \\ \hline
$0$ & $0.50773226$ & $0.48833347$ & $0.50949336$ \\ \hline
$1$ & $1.45551369$ & $1.44451856$ & $1.45972923$ \\ \hline
$2$ & $2.36941282$ & $2.36286881$ & $2.37461896$ \\ \hline
$3$ & $3.25544187$ & $3.25137213$ & $3.26106459$ \\ \hline
$4$ & $4.13235379$ & $4.12882619$ & $4.13805287$ \\ \hline
$5$ & $4.95997506$ & $4.96305356$ & $4.96478901$ \\ \hline
\end{tabular}%
\caption[The eigenvalues of quantum dot]{The eigenvalues Hamiltonians $%
H_{2}, $ $H_{3}$ and $H_{4},$ for $\protect\gamma =0.1.$ The result is
obtained after $30$ iterations.}
\end{table}

The results given in Table II shows that eigenvalues and eigenfunctions are
also depends on the choices of the parameters, $\varepsilon ,\rho ,$ and $%
\eta $ of Hamiltonian (\ref{eq:1}).

\section{Remarks and Discussions}

In this paper, we have studied the solution of the PDM Schr\"{o}dinger
equation without mapping the potential in to the mass space. We have solved
PDM Schr\"{o}dinger equation for four different kinetic energy operators
including harmonic oscillator potential with the variable mass function of
the form $m(x)=m_{0}\left( 1+\gamma x^{2}\right) $. It is shown that energy
levels of the PDM Schr\"{o}dinger equation depends on the mass
distributions. It is important to remark that the results presented here,
shows that eigenvalues also depends on the ordering parameters of the PDM
Schr\"{o}dinger equation \cite{Dutra3}.

We have presented an approximate method based on asymptotic Taylor Series
Expansion of a function. Fortunately, this method is useful for obtaining
both eigenvalues and eigenfunctions of the Schr\"{o}dinger type equations.
Therefore, the results have been obtained here, allowing further comparisons
between the models.

As a further work the method presented here can be used to built more
realistic models for the PDM physical systems. Before ending this work a
remark is in order. When the potential mapped to the mass space, the both
constant and PDM Hamiltonian has the same eigenvalues. It will be worthwhile
to discuss physical acceptability of such an isospectrality in the position
dependent mass background. Therefore we have to develop methods for solving
PDM Schr\"{o}dinger equation without connecting mass to potential or vice
versa.

\section{Acknowledgement}

The research was supported by the Scientific and Technological Research
Council of \ TURKEY (T\"{U}B\.{I}TAK).


\section{References}

\begin{thebibliography}{99}
\bibitem{Serra} Serra L I and Lipparini E 1997 EPL 40 667

\bibitem{Barranco} Barranco M, Pi M, Gatica S M, Hernandez E S, and Navarro
J 1997 Phys. Rev. B 56 8997

\bibitem{Bastard} Bastard G 1988 \textit{Wave Mechanics Applied to
Semiconductor Heterostructure} (Editions de Physique, Les Ulis)

\bibitem{Koc0} Koc R, Koca M, Sahinoglu G 2005 Eur. Phys. J B 48 583

\bibitem{Geller} Geller M R and Kohn W 1993 Phys. Rev. Lett. 70 3103.

\bibitem{Arias} Arias de Saavedra F, Boronat J, Polls A and Fabrocini A 1994
Phys. Rev. B 50 4248

\bibitem{Harrison} Harrison P 2000 \textit{Quantum Wells, Wires and Dots}
(New York: Wiley)

\bibitem{Weis} Weisbuch C and Vinter B 1993 \textit{Quantum Semiconductor
Heterostructure} (New York: Academic)

\bibitem{Peter} Peter A J and Navaneethakrishnan K 2008 Physica E 40 2747

\bibitem{Khor} Khordad R 2010 Physica E 42 1503

\bibitem{Alex} Schmidt A G M, Azeredo A D and Gusso A 2008 Phys. Lett. A 372
2774

\bibitem{Taro} Ando T and Ohtake Y 2006 Pyhs. Rev. E 73 066702

\bibitem{Fu} Fu Y and Chao K A 1989 Phys. Rev. B 40 8349

\bibitem{Young} Young K 1989 Phys. Rev. B 39 13434

\bibitem{von1} von Roos O and Mavromatis H 1985 Phys. Rev. B 31 2294

\bibitem{Haidari} Alhaidari A D 2004 Phys. Lett. A 322 72

\bibitem{Roy} Roy B 2006 Mod. Phys. Lett. B 20 1033

\bibitem{Ikdar1} Ikhdair S M 2010 J. Math. Phys. 51 023525

\bibitem{Peng} Peng X L, Liu J Y and Jia C S 2006 Phys. Lett. A 352 478

\bibitem{Ikdar2} Ikhdair S M 2009 Eur. Phys. J A 40 143

\bibitem{Arda} Arda A, Sever R and Tezcan C 2009 Phys. Scr. 79 015006

\bibitem{Dutra} Dutra A D and Jia CS 2006 Phys. Lett. A 352 484

\bibitem{Quesne1} Quesne C. 2009 SIGMA 5 046

\bibitem{Tezcan} Tezcan C and Sever R 2007\ J. Math. Chem. 42 387

\bibitem{Chen} Chen Gang 2005 Chinese Phys. 14 460

\bibitem{Mustafa} Mustafa O and Mazharimousavi S H 2006 J. Phys. A 39 10537

\bibitem{Roy2} Roy B 2005 EPL 72 1

\bibitem{Kerimov} G A Kerimov 2009 J. Phys. A: Math. Theor. 42 445210

\bibitem{Yahio} Yahiaoui S A and Bentaiba M 2009 Int. J. Theor. Phys. 48 315

\bibitem{Quesne2} Quesne C, 2007 J. Phys. A 40 13107

\bibitem{Dong} Dong S H, Pena J J, Pacheco-Garcia C and Garcia-Ravelo J 2007
Mod. Phys. Lett. A 22 1039

\bibitem{Roy3} Roy B and Roy P 2005 Phys. Lett. A 340 70

\bibitem{Jana} Jana T K and Roy P 2009 EPL 87 30003

\bibitem{Cruz} Sara Cruz y Cruz and Oscar Rosas-Ortiz 2009 J. Phys. A: Math.
Theor. 42 185205

\bibitem{Quesne3} Quesne C and Tkachuk V M 2004 J. Phys. A: Math. Gen. 37
4267

\bibitem{Tanaka} Tanaka T 2006 J. Phys. A: Math. Gen. 39 219

\bibitem{Suzko} Suzko A A and Schulze-Halberg A 2008 Phys. Lett. A 372 5865

\bibitem{Koc1} Koc R and Tutunculer H\ 2003 Ann. Phys. (Leipzig) 12 684

\bibitem{Bagchi} Bagchi B and Tanaka T 2008 Phys. Lett. A 372 5390

\bibitem{Ganguly} Ganguly A and Nieto L M 2007 J. Phys. A: Math. Theor. 40
7265

\bibitem{Gonul} Gonul B, Ozer O, Gonul B and Uzgun F 2002 Mod. Phys. Lett A
17 2453

\bibitem{Bagchi1} Bagchi B, Banerjee A, Quesne C, Tkachuk V M 2005 J. Phys.
A: Math. Gen. 38 2929

\bibitem{Cannata} Cannata F, Loffe M V and Nishnianidze D N 2008 Ann.
Phys.(NY) 323 2624

\bibitem{Roy4} Roy B and Roy P J.Phys.A:Math. Gen 35 (2002) 3961.

\bibitem{Koc2} Ko\c{c} R and Koca M, J. Phys. A: Math. Gen. 36 (2003) 8105.

\bibitem{Koc3} Ko\c{c} R, Koca M and K\"{o}rc\"{u}k E, J. Phys. A: Math.
Gen. 35 (2002) L527.

\bibitem{Gang} Gang C 2004 Phys. Lett. A 329 22

\bibitem{Chen2} Chen G, Chen ZD 2004 Phys. Lett. A 331 312

\bibitem{Bagchi2} Bagchi B, Gorain P, Quesne C and Roychoudhury R 2004 Mod.
Phys. Lett. A 19 2765

\bibitem{Haidari2} Alhaidari A D 2003 Int. J. Theor. Phys. 42 2999

\bibitem{Jiang} Jiang L, Yi L Z and Jia C S 2005 Phys. Lett. A 345 279

\bibitem{Ioffe} Ganguly A, Ioffe M V and Nieto L M 2006 J. Phys. A: Math.
Gen. 39 14659

\bibitem{Kuru} Ganguly A, Kuru S, Negro J and Nieto, LM 2006 Phys. Lett. A
360 228

\bibitem{Carin} Carinena J F, Perelomov A M, Ranada MF and Santander M 2008
J. Phys. A: Math. Theor. 41 085301

\bibitem{Midya} Midya B and Roy B 2009 Phys. Lett. A 373 4117

\bibitem{Qu} Ou Y C, Cao Z Q and Shen Q H 2004 J. Phys. A: Math. Gen. 37 4283

\bibitem{Chen3} Zi-Dong Chen and Gang Chen 2005 Phys. Scr. 72 11

\bibitem{Gorain} Bagchi B, Gorain P, Quesne C and Roychoudhury R 2005 Eur.
Phys. Lett. 72 155

\bibitem{Ju} Ju Guo-Xing et al 2007 Commun. Theor. Phys. 47 1001

\bibitem{Aktas} Aktas M and Sever R 2008 J. Math. Chem. 43 92

\bibitem{Sever} Sever R, Tezcan C, Yesiltas O and Bucurgat M 2008 Int. J.
Theor. Phys. 47 2243

\bibitem{Kraen} Kraenkel R A and Senthilvelan M 2009 J. Phys. A: Math.
Theor. 42 415303

\bibitem{Dutta} D. Dutta and P. Roy 2010 EPL 89 20007

\bibitem{Koc4} Ko\c{c} R and Ol\u{g}ar E, arXiv:1008.0697.

\bibitem{Ciftci} \c{C}ift\c{c}i H, Hall R L and Saad N 2003 J. Phys. A:
Math. Gen\textit{.} \textbf{36} 11807.

\bibitem{Taylor} Struik D\ J 1969 \textit{A Source Book in Mathematics (}%
Translated into English) (Cambridge, Massachusetts: Harvard University Press)

\bibitem{von2} von Roos O 1993 Phys. Rev. B 27 7547.

\bibitem{Dutra2} Dutra A D 2006 J. Phys. A 39 203

\bibitem{Dekar} Dekar L, Chetouani L and Hammann T F 1998 J. Math. Phys. 39
2551.

\bibitem{ciftci} \c{C}ift\c{c}i H, Hall R L and Saad N 2003 \textit{J. Phys.
A: Math. Gen.} \textbf{36} 11807

\bibitem{boyd} Boyd J P 1999 Acta Applicandae Mathematicae 56 1

\bibitem{Dutra3} Dutra A D and Oliveira J A Phys. Scr. 78 (2008) 035009
\end{thebibliography}
\end{document}